\documentclass[aps,prl,groupedaddress,superscriptaddress,twocolumn,10pt,longbibliography]{revtex4-2}
\pdfoutput=1 
\usepackage[utf8]{inputenc}
\usepackage{hyperref}
\usepackage{graphicx}
\usepackage{setspace}
\usepackage{orcidlink}
\usepackage{soul}

\newcommand{\cmtout}[1]{}
\newcommand{\kc}{k_{SP}}

\usepackage{siunitx} 
\newcommand{\omatt}[1]{}

\usepackage[caption=false]{subfig}
\usepackage{tabularray}
\usepackage{amssymb}
\usepackage{physics}
\usepackage{float}
\DeclareSIUnit\bohr{\text{\ensuremath{a_\textup{0}}}}

\begin{document}


\title{Vibrationally highly excited trilobite molecules stabilized by non-adiabatic coupling}

\author{Rohan Srikumar\orcidlink{0000-0003-0303-1331}}
\thanks{These authors contributed equally to this work.}
\affiliation{Zentrum für Optische Quantentechnologien, Universität Hamburg, Luruper Chaussee 149, 22761 Hamburg, Germany}

\author{Markus Exner\orcidlink{0009-0005-8290-7371}}
\thanks{These authors contributed equally to this work.}
\affiliation{Department of Physics and Research Center OPTIMAS, Rheinland-Pfälzische Technische Universität Kaiserslautern-Landau, 67663 Kaiserslautern, Germany}

\author{Richard Blättner\orcidlink{0009-0004-4667-821X}}
\affiliation{Department of Physics and Research Center OPTIMAS, Rheinland-Pfälzische Technische Universität Kaiserslautern-Landau, 67663 Kaiserslautern, Germany}

\author{Peter Schmelcher\orcidlink{0000-0002-2637-0937}}
\affiliation{Zentrum für Optische Quantentechnologien, Universität Hamburg, Luruper Chaussee 149, 22761 Hamburg, Germany}
\affiliation{The Hamburg Centre for Ultrafast Imaging, Universität Hamburg, Luruper Chaussee 149, 22761 Hamburg, Germany}

\author{Matthew T. Eiles\orcidlink{0000-0002-0569-7551}}
\affiliation{Max Planck Institute for the Physics of Complex Systems,  Nöthnitzer Str. 38, 01187 Dresden, Germany}

\author{Herwig Ott\orcidlink{0000-0002-3155-2719}}
\thanks{Corresponding author: ott@physik.uni-kl.de}
\affiliation{Department of Physics and Research Center OPTIMAS, Rheinland-Pfälzische Technische Universität Kaiserslautern-Landau, 67663 Kaiserslautern, Germany}

\date{\today}

\begin{abstract}
We report on the observation of highly excited ($\nu \sim 100)$ vibrational states of a trilobite ultralong-range Rydberg molecule in $^{87}$Rb. 
These states manifest spectroscopically in a regularly spaced series of peaks red-detuned from the $25f_{7/2}$ dissociation threshold. 
The existence and observed stability of these states requires the almost complete suppression of the adiabatic decay pathway induced by the $P$-wave shape resonance of Rb. 
This stabilization is predicted to occur only for certain Rydberg levels where the avoided crossing between trilobite and $P$-wave dominated butterfly potential energy curves nearly vanishes, allowing the vibrational states to diabatically traverse the crossing with almost unit probability.  
This is the first direct measurement of beyond-Born-Oppenheimer physics in long-range Rydberg molecules, and paves the way for future experiments to access and manipulate wavepackets formed from high-lying vibrational states.


\end{abstract}

\maketitle
Efforts to understand and harness ultracold molecules including Rydberg atoms among their constituents are essential to the study of impurity dynamics in ultracold gases \cite{Schmidt_2016,Sous_2020,tiwari2022dynamics,Camargo_2018,Durst_2024,Engel2024,Greene2016,Pfau2008,kleinbach2018ionic,berngruber2024situ}, the design of quantum information storage systems \cite{Hofferbirth2017}, precision measurements of electron-atom interactions \cite{Deiglmayer2015,Engel_2019,Giannakeas2020,Raithel_2019,Deiglmayer2021}, and the wider study of unorthodox molecular physics and ultracold chemistry \cite{Deiss2020,Geppert2021,Pfau2016,Duspayev_2021,Deiss_2021,Ott_2015,Zuber2022,anasuri_2023}. 
In particular, ultralong-range Rydberg molecules (ULRMs) composed of a Rydberg atom and a ground-state atom have provoked significant interest due to the possibility that the ground-state atom hybridizes the degenerate Rydberg states of high angular momentum $\ell$ \cite{Greene2000,Hummel2020,Eiles_2019,Dunning_2024}. 
The resulting molecular states, dubbed trilobites and butterflies depending on the angular momentum ($S$ or $P$-wave, respectively) of the Rydberg electron with respect to the ground-state atom, 
are highly desirable as they possess enormous dipole moments \cite{Niederprum2016,Booth_2015,Althoen2023} and are more deeply bound than their non-polar counterparts. 
The flexibility of the high-$\ell$ states to adapt to the nuclear geometry makes these states also exciting to investigate in the context of polyatomic systems, ranging from triatomic molecules to Rydberg composites with electronic behavior analogous to solid-state systems \cite{liu2009ultra,Fey2019,Eiles_2020,eiles2020ring,Eilesanderson,Eilestopology,Eisfeld2020,Hummel2019}.
\begin{figure*}[t]
\centering
\includegraphics[width=1\textwidth]{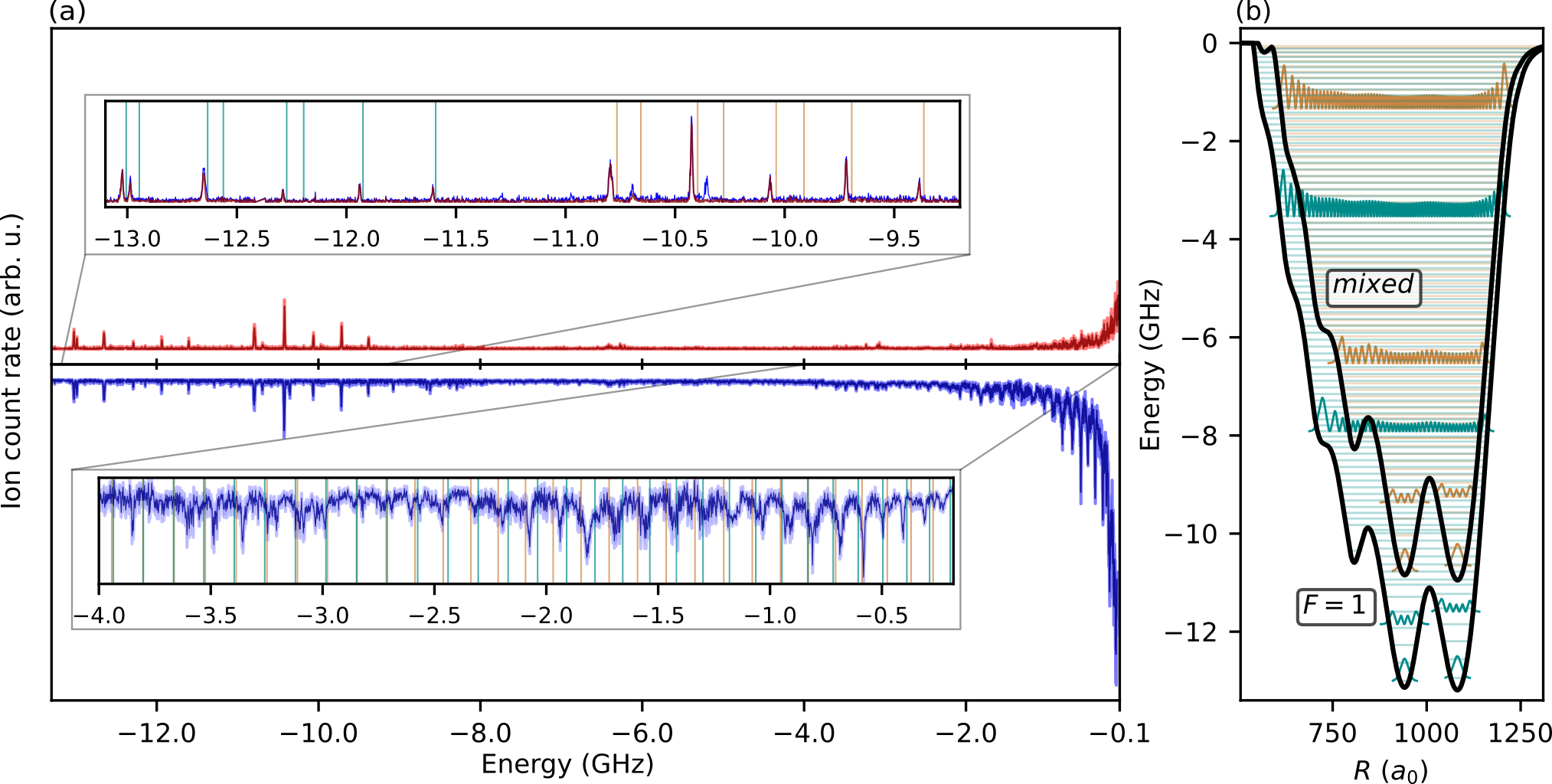}
\caption{(a) Ion signal as a function of the detuning from the $^{87}$Rb 25$f_{7/2}$ atomic state. In red (blue) is the signal of ions that had zero (finite) momentum before ionization. The light-blue (-red) colored region around the finite (zero) momentum spectra indicates the ion count rate error. The upper inset shows a magnification of the energy range containing the deeply bound trilobite states with the theoretical spectrum overlaid as a stick plot. 
The lower inset shows a magnification of the spectrum of highly excited vibrational states. The peak heights are scaled linearly in energy to improve visibility. The colored lines mark the positions of the theoretical levels, bound in the two diabatic trilobite PECs shown in (b). In panel (b), selected vibrational wave functions are also plotted. }
\label{fig:25_spectrum}
\end{figure*}

Exploiting ULRMs for these promising applications has been hindered by the difficulty of exciting trilobite molecules. 
Their high-$\ell$ nature implies a weak dipole coupling to intermediate states, which is overcome either by mixing these states with nonpolar low-$\ell$ Rydberg states, thus reducing their advantageous properties, or via a three-photon excitation scheme.
These approaches have led to the successful observation of low-lying vibrational trilobite states \cite{Althoen2023,Exner2024,Booth_2015,Kleinbach_2017}. 
However, a significant impediment to the creation of vibrationally excited trilobite states is the existence of scattering resonances in the $P$-wave channel.  
These introduce avoided crossings between the trilobite potential energy curve (PEC), which possesses a global minima, and the butterfly PEC, which is in this energy range steeply attractive without any stable minima \cite{Hamilton_2002,Chibisov_2002}.
In turn, the inner repulsive wall of the trilobite PECs are destroyed, suppressing higher vibrational excitations.
Although theoretical calculations have demonstrated the ability of non-adiabatic coupling to modify the binding energies and stability of these states to some extent, no unambigious experimental confirmation yet exists for any beyond Born-Oppenheimer effects in ULRM vibrational states \cite{Srikumar2023,Hummel_2022,durst2024}.

In this Letter, we report the experimental detection of highly excited vibrational states of a $^{87}$Rb trilobite molecule which are red-detuned from the $25f_{7/2}$ dissociation threshold.
By comparison with the calculated spectrum, we assign vibrational quantum numbers $\nu \gtrsim 80$ to these levels. This assignment only succeeds by treating the trilobite PEC alone, totally neglecting the butterfly interaction.
This assumption, consistent with a single-channel \textit{diabatic} approach, is motivated by a recent theoretical model developed to understand the behavior of avoided level crossings in trilobite and butterfly PECs ~\cite{eiles2024}. Furthermore, this model predicts that the $n=27$ PECs behave fully adiabiatically, which we confirm experimentally by performing spectroscopy below the $27f_{7/2}$ threshold and finding no signatures of high-lying vibrational states.  
Our findings therefore provide the first direct experimental evidence for strong non-adiabatic coupling in Rydberg molecules.

Fig.~\ref{fig:25_spectrum}(a) shows the complete vibrational spectrum of a long-range trilobite Rydberg molecule over an energy range stretching from the $25f_{7/2}$ atomic Rydberg line at zero detuning to the bottom of the trilobite potential energy curve at approximately \SI{-13}{GHz} \cite{Exner2024}.
This spectrum was obtained from an atomic sample of $^{87}$Rb atoms at \SI{40}{\micro K} and with a peak density of \SI{4 e13}{cm^{-3}}. 
The atoms are prepared in the $F=1$ hyperfine ground state in a crossed dipole trap at $\lambda = \SI{1064}{nm}$. 
We excite Rydberg $nf$ states through a three photon excitation scheme at wavelengths of \SI{780}{nm}, \SI{776}{nm}, and \SIrange{1276}{1288}{nm}. 
Subsequently, we ionize any Rydberg atoms produced and detect the ions in a reaction microscope.
This allows us to spectroscopically distinguish long-lived molecules having zero momentum prior to ionization (red signal) from short-lived states whose decay imparted some finite momentum (blue signal) \footnote{see \cite{Exner2024,Althoen2023} and the End Matter section E1 for details}.

The spectral peaks with binding energies deeper than \SI{-9}{GHz}, which have already been reported in Ref.~\cite{Exner2024}, are the signatures of polar trilobite molecules with vibrational states bound in local minima in the deep trilobite potential. 
We have assigned vibrational quantum numbers up to $\nu=5$ in the various potential wells to these peaks \cite{Exner2024}. 
Higher in energy, the signal strength declines rapidly, esentially vanishing from \SI{-8}{GHz} to \SI{-4}{GHz}.
The signal resumes from $-4$~GHz, growing rapidly in strength until the atomic resonance. 
It is primarily detected in the full momentum spectra, indicating that these states are more short-lived than the deeply bound states. 
This series is highlighted in the bottom inset, where the peak heights are scaled linearly with energy for easier visualization. 
Upwards of thirty peaks with a consistent spacing of \SI{\sim 120}{MHz} are observed.

\begin{figure}[h]
\centering
\includegraphics[scale=1]{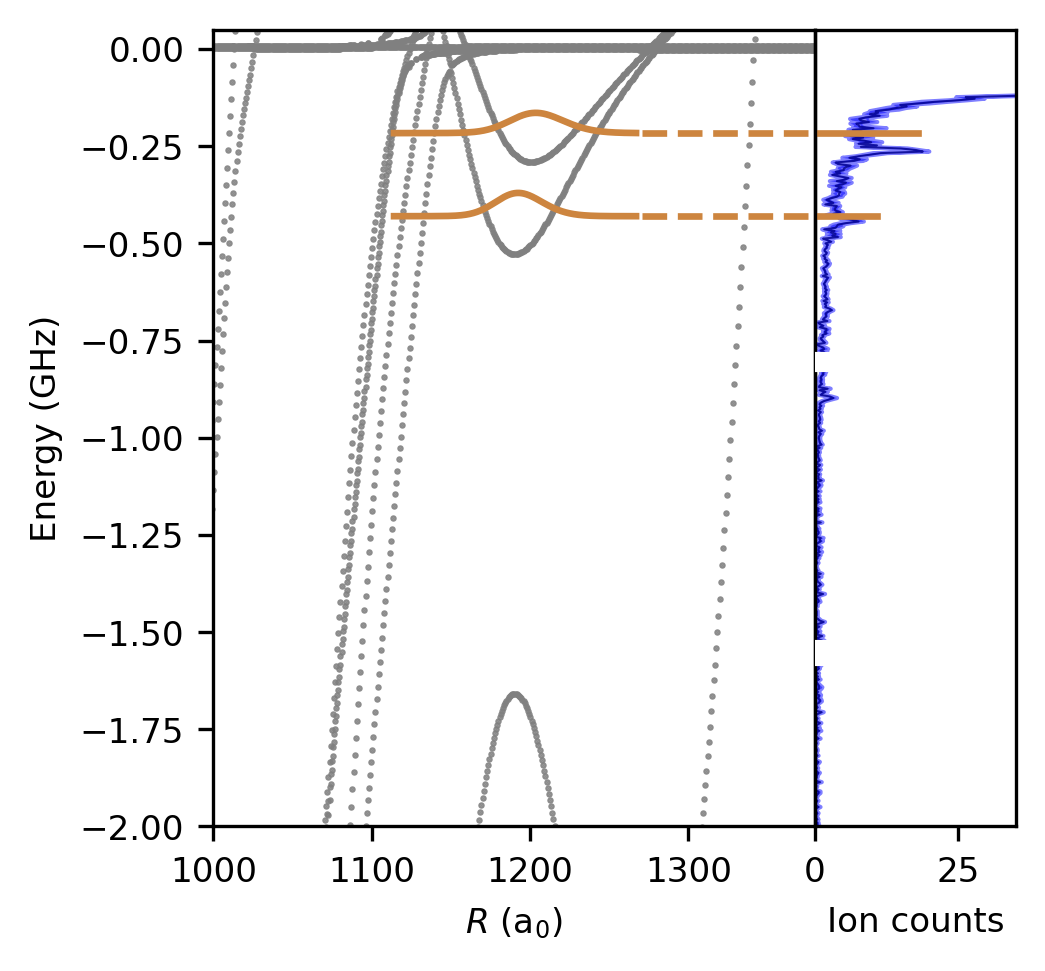}
\caption{The molecular spectra and associated potential energy curves near the 27$f_{7/2}$ dissociation threshold. The potential energy curves (grey) possess shallow wells which support localized vibrational states (dark-orange) in good agreement with the experiment (blue spectrum). The two vibrational states calculated are shown in the orange stick spectrum, with their heights scaled by their respective Franck-Condon factors. } 
\label{fig:near_resonant}
\end{figure}

Spectroscopic assignment of these peaks  is accomplished by comparison of this spectrum with that obtained from theoretical calculations. 
As in Ref.~\cite{Exner2024}, we have used an approach based on the Coulomb Green's function and incorporating the fine and hyperfine structure of both Rb atoms and all spin-dependent $e-$Rb scattering channels to compute the PECs \cite{eiles2023}. 
Within the Born-Oppenheimer approximation, the adiabatic potential energy curves from this calculation are not repulsive at short-range; they instead drop steeply due to the effect of the $P$-wave interaction. 
Such a potential can still support excited internal reflection states, reflected off of the steep drop \cite{Bendkowsky2009}, but the predicted level spacing ($\sim 230$MHz) and resonance widths ($\sim$ 260MHz) are wholly incompatible with the measured spectrum. 

However, the vibrational spectrum of the purely \textit{diabatic} trilobite PECs shown in Fig.~\ref{fig:25_spectrum}(b) is in very good agreement with the observed series (see the stick spectrum overlaid in the inset of panel (a)). 
From this calculation, we identify these peaks as the photoassociation signal of highly excited vibrational states with $\nu$ in excess of 80.

Remarkably, the opposite is true for $n=27$, as shown in Fig.~\ref{fig:near_resonant}. The long series of almost equidistant molecular resonances observed for $n=25$ is completely absent from the experimental spectra for $n=27$. 
Only two vibrational peaks are detected. The adiabatic PECs support two quasi-bound molecular resonances in this same energy range, and their binding energies and line strengths are in good agreement with those observed.

\begin{figure*}[t]
\centering
\includegraphics[scale=1]{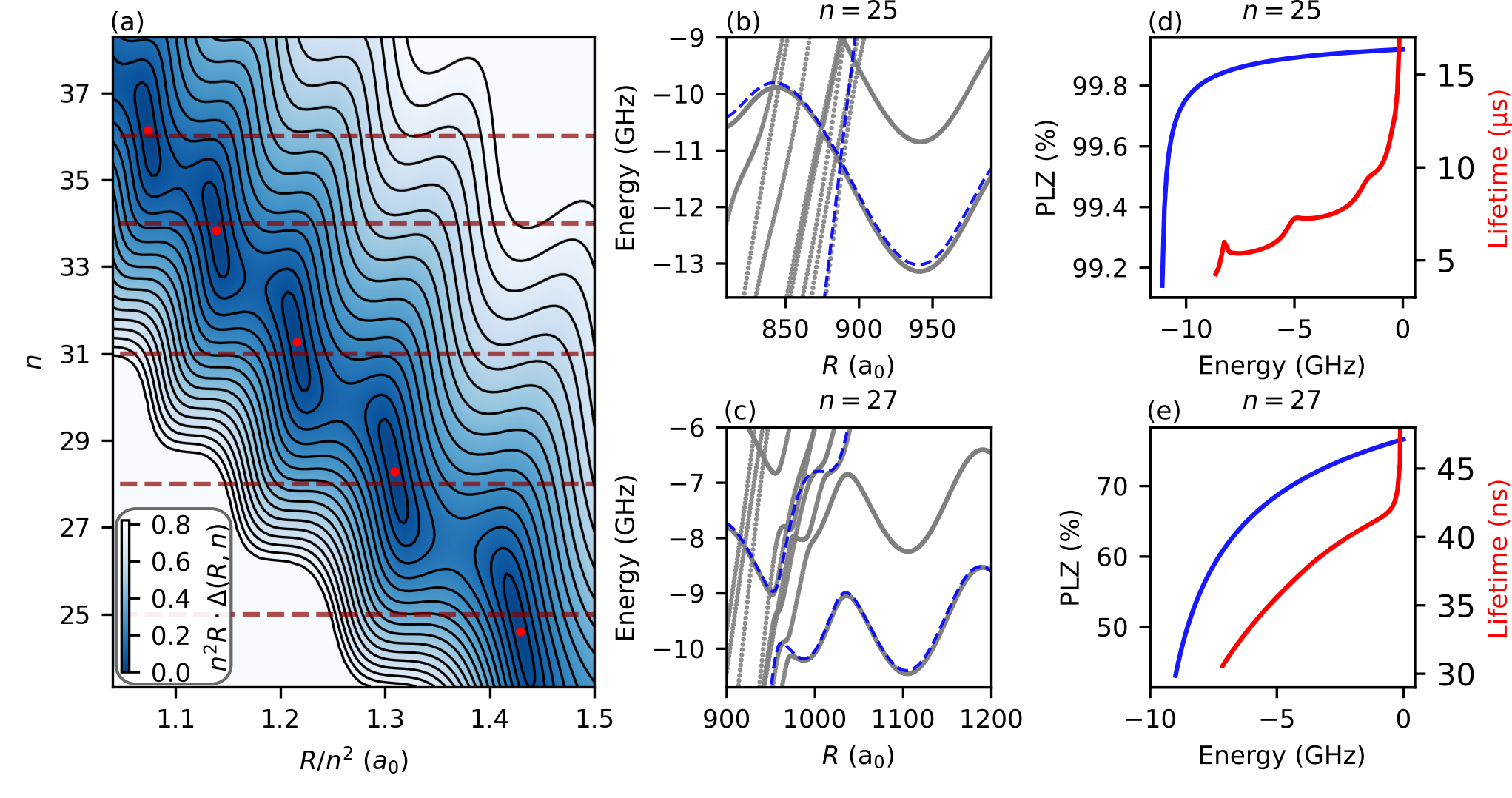}
\caption{ A contour plot of $n^2  R \cdot \Delta(R,n)$ shows the existence of regular intersections in the potential energy surfaces $V_\pm(R,n)$. At the dark blue regions, $\Delta(R,n)\sim 0$, and it exactly vanishes at the marked points where Eq.~\ref{eq:ucondition} is satisfied. The horizontal lines highlight $n$ levels which should display a high degree of diabaticity. 
The adiabatic potential curves calculated via the spin-dependent Green's function method (grey solid-line) and the effective two-level model (dashed blue-line) are shown for $n$=25 (b), and $n=27$ (c).
As predicted, the potentials are far more diabatic in the $n=25$ case, where the avoided crossings nearly vanish.
The Landau-Zener crossing probability $P_{LZ}$ (red) between the two adiabatic states, and the associated Landau-Zener lifetimes (blue) are shown for $n$=25 (d) and $n$=27 (e). }
\label{fig:contour}
\end{figure*}

\begin{figure}[t]
\centering
\includegraphics[scale=1]{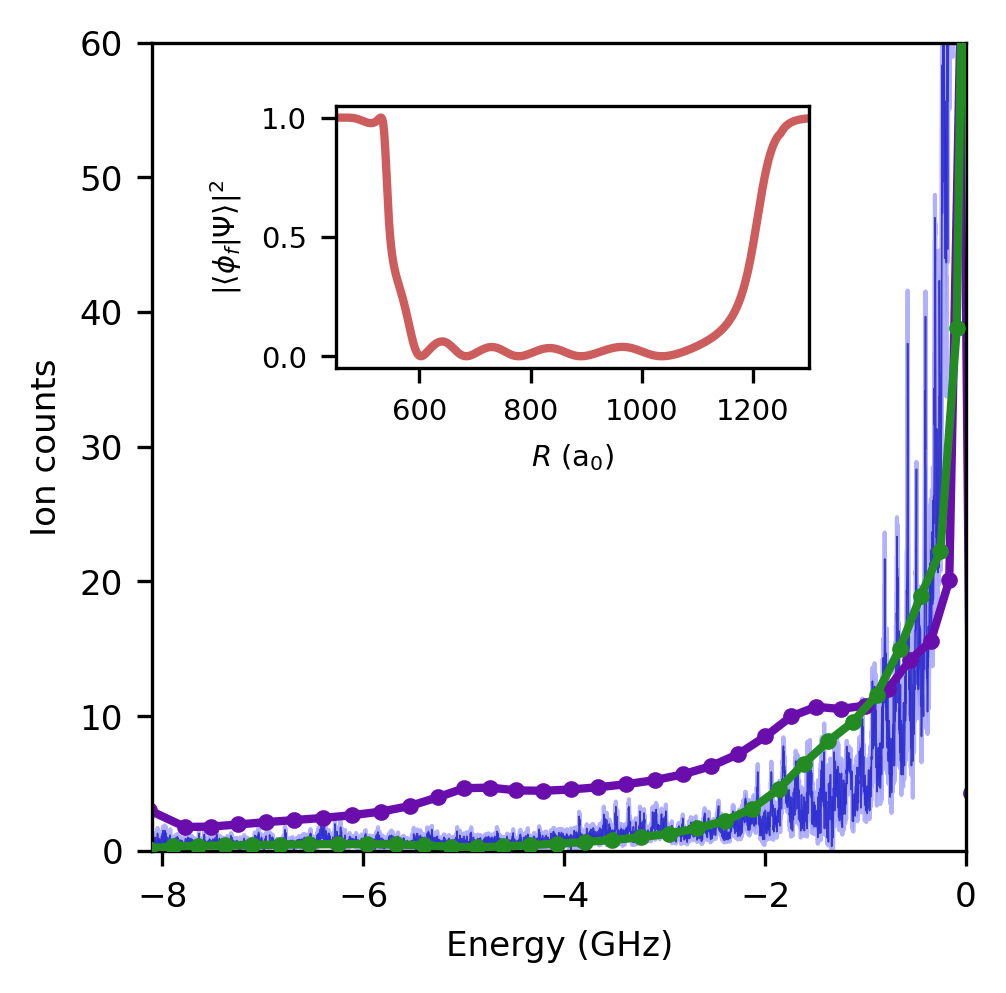}
\caption{Energy-dependent Franck-Condon factor calculated neglecting (purple) and including (green) the $f$ state admixture $|\langle \phi_f|\Psi(R)\rangle|^2$ shown in the inset. Both curves are scaled so that their values are equal at $E$=0.
}
\label{fig4}
\end{figure}

 Ref.~\cite{eiles2024} presented a spin-independent model of the trilobite and butterfly molecular system which predicts this observed behavior. 
In this model, dynamics in the  molecular system follows the two adiabatic potential energy {surfaces} 
\begin{align}
\label{eq:vsurfaces}
    V_\pm(R,n) &= \frac{U_+(R,n)}{2}\pm \frac{\Delta(R,n)}{2}
\end{align}
in the synthetic coordinate space spanned by the internuclear distance $R$ and the principal quantum number $n$, taken to be a \textit{continuous} variable. 
Here, $
U_\pm(R,n)=U_S(R,n) \pm U_P(R,n)$
are superpositions of the pure $S$-wave and pure $P$-wave dominated PECs \footnote{See End Matter section E2 for details}, and 
\begin{align}
\label{eq:delta}
    \Delta(R,n)&=\Bigg[\left[U_-(R,n)\right]^2\\&\nonumber+\sqrt{\frac{12\pi \tan\delta_S(k)\tan\delta_P(k)}{k^{4}}}\left(\frac{u_{n0}(R)}{4\pi R}\right)^4\Bigg]^{1/2}
\end{align}
is the energy gap between adiabatic potential energy surfaces. 
\cmtout{\begin{align}
    \mathcal{C}(R,n)&=\sqrt{\frac{12\pi \tan\delta_S(k)\tan\delta_P(k)}{k^{4}}}\left(\frac{u_{n0}(R)}{4\pi R}\right)^4
\end{align} }
The energy-dependent scattering phase shifts are $\delta_S(k)$ and $\delta_P(k)$, and $u_{n0}(R)$ is the reduced $s$-state radial wave function, taken to be the regular Whittaker solution so that it remains valid for arbitrary non-integral $n$. 

A useful characteristic for the degree of adiabaticity is the Landau-Zener probability for a non-adiabatic transition,
\begin{equation}
    P_\mathrm{LZ} =  \exp\left(-2\pi\Delta^2/4vs\right),
\end{equation}
 where $s = |d\Delta/dR|$ is the differential slope at the crossing and $v=\sqrt{2\mu E - \mu(V_+ + V_-)}$ is the semiclassical velocity of the vibrational state.
Generally speaking, a purely diabatic treatment requires  $P_\mathrm{LZ}\to 1$, which is implied by the limit $\Delta\to 0$. We therefore seek the conditions which enable the gap to close. 
Both terms in Eq.~\ref{eq:delta} are non-negative, so the gap closes only when both terms simultaneously vanish. As shown in Ref.~\cite{eiles2024}, an application of the spatially generalized Kato's theorem \cite{Kato1957,March1986} leads to the system of equations 
 \begin{align}
 \label{eq:ucondition}
     u_{n0}(R) = 0\,\,\text{and}\,\,
    R = \frac{2 n^2}{1+(\kc  n)^2};
\end{align}
where $\kc^2/2\approx 9$meV is the energy where the $S$ and spin orbit-averaged $P$-wave scattering phase shifts are identical \footnote{See End Matter section E2 for more details}.

Fig.~\ref{fig:contour} shows a contour plot of $\Delta(R,n)$ scaled by $n^2 R$ for ease of visualization.
The marked points indicate where Eq.~\ref{eq:ucondition} is satisfied.
A Rydberg state whose integer-valued principal quantum number lies close to a one of these points will exhibit a strong degree of diabaticity. 
The horizontal lines show that $n=25$, along with several higher $n$ values, comes very close to satisfying these conditions, while $n=27$ certainly does not. 
In Fig.~\ref{fig:contour}(b) and (c), we compare the adiabatic PECs in this model (Eq.~\ref{eq:vsurfaces}, dashed blue lines) with those obtained in the full spin-dependent approach (solid gray) for $n=$25 and $n=$27 respectively. 
These are in excellent agreement regarding both the size of the avoided crossing and the overall depth and structure of the potential curves, confirming the accuracy of this two-level treatment. 

Figure \ref{fig:contour} shows the calculated $P_{LZ}$ (blue) and lifetimes (red) as a function of energy for $n$=25 (d) and $n$=27 (e).
As expected, the former case is extremely diabatic with $P_{LZ}\sim 1$ almost independent of the velocity.
We approximate the lifetime of a molecular state with binding energy $E$ as
$\tau (E) = T/(1-P_{LZ})$, where $T$ is the oscillation period of the vibrational state \footnote{see End Matter section E3 for more details}. 
For $n=25$ these are on the order of a few microseconds, comparable with the radiative lifetime.
As this approach includes the effect of only a single crossing out of the many seen in Fig.~\ref{fig:contour}, this estimate should be taken as an upper bound on the lifetime \footnote{see End Matter section E4 for more details.}.
For $n$=27 the hopping probability is strongly energy-dependent, as now $\Delta$ is large enough that the contribution of $v$ in the exponential is important. 
It does not exceed 60\% even for the highest-lying states.
The calculated lifetimes are correspondingly short, on the order of a few tens of nanoseonds. 

Finally, we investigate the observed decline of the signal strength as a function of detuning.  
Such a decline cannot be based on a change in Landau-Zener probability between states, due to its insensitivity to the semi-classical velocity. 
We compute the Franck-Condon factor (FCF) $f_\nu=\int \psi_\nu d(R)\psi_i dR$ of each vibrational state $\psi_\nu$.
Here $\psi_i$ is the initial scattering state of the two atoms and $d(R)$ is the $R$-dependent electronic dipole moment. 
This quantity, proportional to the $f$-state admixture in the electronic wave function, is shown in the inset of Figure \ref{fig4}. Both $d(R)$ and the vibrational states are calculated within a model containing only the diabatic trilobite potential curve and the $f$-state PEC.  

Figure \ref{fig4} shows the energy dependence of the FCF (blue curve) by connecting its discrete points across the vibrational spectrum. 
Its behavior is in qualitative agreement with the peak heights obtained experimentally. 
The comparison between this curve and the curve plotted in orange, which does not include the transition dipole moment $d(R)$, illustrates how the energy dependence  of the FCF is strongly modified by the rapid increase in $f$-state admixture on the edges of the trilobite well ($R>1200$ and $R<600$). 
As $\nu$ increases, the vibrational states extend into these regions of $R$ and therefore couple much more strongly to the initial state.

In conclusion, we have reported the first observation of  high-lying vibrational states (up to $\nu = 100$)  of trilobite molecules associated with the $n=25$ Rydberg state. 
Such states are protected from internal molecular decay since the set of trilobite and butterfly PECs is almost perfectly diabatic for this $n$. 
This is not the case in $n=27$, and such states are subsequently not observed experimentally, confirming the prediction of strongly $n$-dependent beyond Born-Oppenheimer effect in \cite{eiles2024}.
The measured peak positions and line strengths are in excellent agreement with theory, further confirming their assignment to highly excited vibrational states. 

Our work reports the first clear evidence of the breakdown of Born Oppenheimer approximation in the bound-states of ULRMs. 
A further experimental exploration of the other $n$ levels that we have predicted will also behave highly diabatically would provide additional confirmation of the employed theory. 
These results pave the way towards wave packet creation in the trilobite potential wells \cite{srikumar2024}, manipulation of the avoided crossing and dynamics around it with external fields \cite{Keiler_2021}, and further study of non-adiabatic effects in the spectra and dynamics of these molecules. 

\textit{Acknowledgements:} We would like to thank Max Althön for helpful discussions. This project is funded by the German science foundation DFG, project numbers 460443971 and 316211972.  
\newline

\bibliographystyle{apsrev4-2}
\bibliography{references}

\vspace{-2mm}
\section{End Matter}

\subsection{E1: Experimental methodology}
After photoassociation of the molecules in a reaction microscope, we ionize them with a CO$_2$ laser.  The resulting ions are directed via an electric field in Wiley-McLaren configuration to a time- and spatially-resolved multi-channel plate detector. This allows us to measure the momentum directly before ionisation, and hence spectroscopically distinguish the molecules based on their lifetime with respect to internal molecular decay via state-changing collisions. These collisions impart larger momenta to the molecular states, and hence a distinction is made between the signal of stable molecules detected at zero momentum (red) and that of shorter-lived molecules detected with finite momentum (blue) in the spectrum shown in Fig. \ref{fig:25_spectrum}. The finite momentum spectrum contains the zero momentum signal, which implies that a resonance peak uniquely detected in the finite momentum signal, and absent in the zero momentum signal, corresponds to a short-lived  molecular resonance (compared to the excitation pulse duration of \SI{3}{\micro\second}).

\subsection{E2: Two-level model}

\begin{figure}[t]
\centering
\includegraphics[width=\columnwidth]{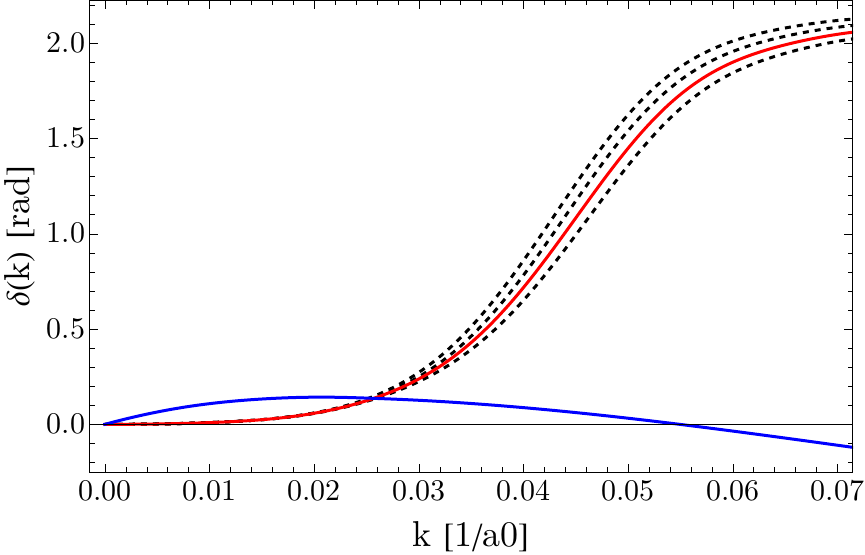}
\caption{ Phase shifts used in this paper. The $^3S_1$ phase shift (blue) intersects the spin-orbit averaged $^3P$ phase shift (red, with black-dashed showing the $^3P_J$ phases) at $k_{SP}=0.0257$}
\label{fig:sm_phases}
\end{figure}

In this section, we provide additional context and calculations for the diabaticity conditions and modeling used in the main text. 
We begin with Fig.~\ref{fig:sm_phases}, which shows the $^3S_1$ (blue) and $^3P_J$ (dashed black) electron-Rb scattering phase shifts across the range of electron momenta relevant here. 
To keep the diabaticity condition simplest, we consider the $^3P$ phase shift (red) which is the spin orbit-averaged result. 
The $S$ and $P$-wave phase shifts are identical at one $k$ value in this energy range, $\kc=0.0257$.

Next, we compute the trilobite and butterfly potential energy curves for the Rb$_2$ long-range Rydberg molecule, setting the $f$-state quantum defect to zero for simplicity and ignoring coupling to the much deeper quantum defect states with $\ell\le 2$. As shown in Ref.~\cite{eiles2024,Eiles_2019}, these potential energy curves are obtained as the eigenvalues of the matrix
\begin{align}
\label{eq:hmat}
    V = \begin{pmatrix}
        U_S & C \\ C & U_P
    \end{pmatrix},
\end{align}
where $U_S=2\pi a_S(k)\sum|\phi_{nl0}(R)|^2$ is the ``trilobite" potential energy curve and $U_P=6\pi a_P^3(k)\sum\left|\frac{d\phi_{nl0}(R)}{dR}\right|^2$ is the ``butterfly" potential energy curve, and $C = \sqrt{12\pi^2a_s(k)a_p^3(k)}\sum\phi_{nl0}(R) \frac{d}{dR}\phi_{nl0}(R)$ is a coupling element between trilobite and butterfly states.  The sums range over all degenerate $l$ levels $3\le l \le n-1$ and $a_S(k)$ and $a_P^3(k)$ are the $S$-wave scattering length $-\tan\delta_S(k)/k$ and $P$-wave scattering volume $-\tan\delta_P(k)/k^3$, respectively. $U_S$ and $U_P$ are separately obtained by diagonalizing the $S$- and $P$-wave pseudopotentials, 
\begin{align}
    \hat V_S  &= 2\pi a_S(k)\delta^3(\vec r - \vec R)\\
    \hat V_P &= 6\pi a_P^3(k)\nabla\delta^3(\vec r - \vec R)\nabla,
\end{align}
separately within the degenerate manifold of a single Rydberg $n$ level.

The potential energy surfaces are then given by Eq.~\ref{eq:vsurfaces} in the main text. Eq.~\ref{eq:delta} is derived by evaluating the sums in $U_S$, $U_P$, and $C$ in closed form to obtain
\begin{align}
\label{eq:sm:UTdef}U_S(R) &= \frac{2\pi a_S(k)}{4\pi}[k^2u^2+u'^2]\\
\label{eq:sm:UBdef}U_P(R) &= 6\pi a_P^3(k)\left(\frac{k^2}{3}\frac{U_S(R)}{2\pi a_S(k)}-\frac{2uu'-u^2/R}{6\pi R^2}\right)\\
\label{eq:sm:Cdef}C&= -\frac{\sqrt{12 a_S(k)a_P^3(k)}\pi u^2}{4\pi R^2}.
\end{align}
In these expressions, $u \equiv u_{n0}(R)$ is a short-hand for the reduced radial $s$-wave Rydberg wave function, and $u'$ is its derivative with respect to $R$.
To evaluate these summations, it is necessary to extend the sum over $\ell$ to include terms with $0\le \ell \le 2$. 
As these terms have only a very small contribution to the total sum, this approximation (equivalent to neglecting the effect of quantum defects) does not affect the accuracy of our conclusions. 
We now can see the diabaticity requirement (defined as a real or extremely narrow avoided crossing between these potential curves) directly from the Hamiltonian matrix of Eq.~\ref{eq:hmat}. If $C = 0$ simultaneously as $U_S = U_P$, the potential curves will cross. The first condition, from Eq.~\ref{eq:sm:Cdef}, holds at any node of the function $u_{n0}(R)$. 
The second condition can be obtained semiclassically from the Borodin and Kazansky expressions \cite{Borodin_1992}, re-derived through a ``dressed-ion" picture in Ref.~\cite{GiannakeasIon2020}, 
\begin{align}
U_S(R) &\approx -\frac{1}{2(n- \delta_S(k)/\pi)^2}\\
U_P(R) &\approx -\frac{1}{2(n - \delta_P(k)/\pi)^2},
\end{align}
and is evidently $\delta_S(k) = \delta_P(k)$. This same condition can be obtained by inserting $u = 0$ into Eqs.~\ref{eq:sm:UTdef} and \ref{eq:sm:UBdef} and solving for $U_T = U_B$. For the given set of phase-shifts \cite{Exner2024}, the $n$-values that satisfy the aforegiven diabaticity condition will exhibit conical intersections in the ($R$,$n$) manifold where  $\Delta(R,n) = (V_+ - V_-)(R,n) \to0$, as seen in Fig.~\ref{fig:contour}.

\subsection{E3: Landau-Zener decay-rates and lifetimes}

In the main text, we characterize the diabatic effects at the trilobite-butterfly crossing by calculating the Landau-Zener transition probability $P_\mathrm{LZ}$,
which gives us the probability of non-adiabatic transition between the two adiabatic curves at an avoided crossing.
Given that $P_{LZ}$ is the probability that the nuclear wave packet stays in the diabatic trilobite,  $P_{d}$=$1-P_{LZ}$ is the probability that the nuclei decays down the adiabatic passage each time the wave packet encounters the avoided crossing.

The number of crossing attempts made before time $t$ can be approximated as $n$ = $t/T$, where $T$ is the period of oscillation (given by the inverse of vibrational energy spacing), for $t\gg T$.
Hence, the probability that the  nuclei remains in the trilobite after $t$=$nT$ has passed is: $$P(t)=(P_{LZ})^n=(1-P_d)^{t/T}.$$
Writing $P_d$ as $1/x$, we get $$P(t) = (1-1/x)^{xt/xT}=((1 - 1/x)^x)^{(t/Tx)}.$$
In the limit $x\to\infty$, consistent with $P_d\to 0$ (or $P_{LZ}\to1$), we have then $$P(t) = e^{-t/Tx} = e^{-t(1-P_{LZ})/T}.$$ 
Hence, for long time scales ($t\gg T$) and high diabaticity ($P_{LZ}\to1$), the population of the trilobite state decays exponentially with the rate $$\Gamma =(1-P_{LZ})/T,$$ giving us the molecular lifetime: $$\tau=\Gamma^{-1}=T/(1-P_{LZ}).$$

\subsection{E4: Landau-Zener calculations for the spin-dependent Green's function model}

\begin{figure}
    \centering
    \includegraphics[width=\linewidth]{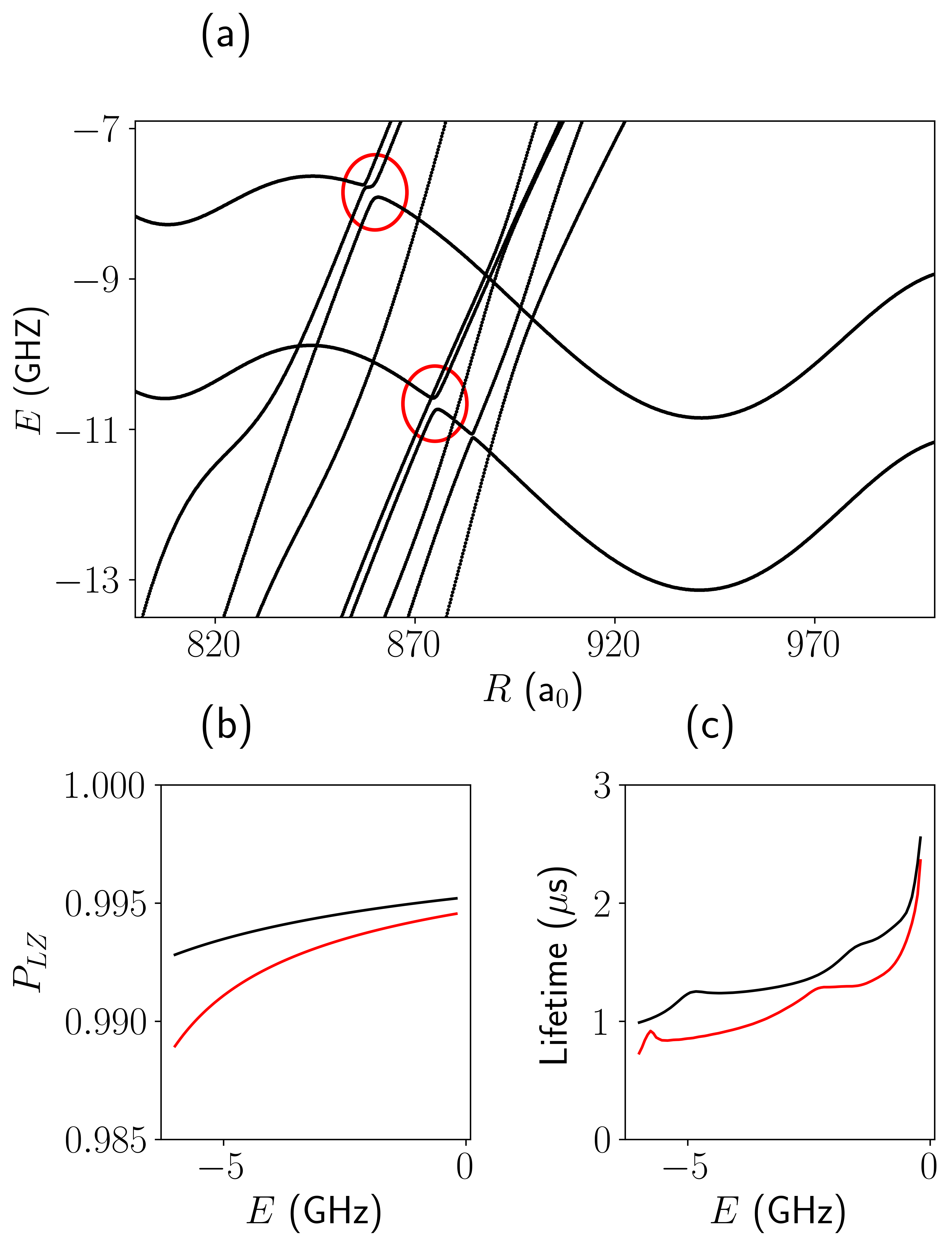}
    \caption{(a) The adiabatic potential energy curves calculated utilizing the spin-dependent Green's function method (black). The avoided crossings that contribute to molecular decay are highighted (red). The Landau-Zener crossing probability (b) and the associated lifetimes (c) are calculated and shown separately for the triplet PECs (black) and the mixed PECs (red).  }
    \label{fig:fullspinLZ}
\end{figure}

The two-level model used to successfully explain the observed stabilization mechanism neglects the full spin-dependent structure of the ULRM \cite{eiles2023,Eiles_2017,Exner2024}. 
The state-of-the-art Green's function model incorporates within its calculations the fine-structure of the Rydberg atom, the hyper-fine structure of the ground-state and and the six different electron-atom scattering channels ($^1S_0,^3S_1,^1P_1,^3P_{0,1,2}$) shown in figure~\ref{fig:sm_phases}.
Figure \ref{fig:fullspinLZ} (a) shows the Born-Oppenheimer potential energy curves of the $n$=25 ULRM in the energy regime  relevant to the avoided crossings. The $^3P_J$ shape resonances introduce multiple crossings between the butterfly and trilobite curves, most of which are near-exact and does not contribute to an adiabatic decay pathway.
However, we do identify one non-negligible avoided crossing between the triplet (mixed) trilobite PEC and the plummeting butterfly curves that presents an adiabatic decay pathway for the molecule in a trilobite state. 
Figure \ref{fig:fullspinLZ} (b) and (c) shows the probability of non-adiabatic transition and the associated lifetimes calculated separately for the triplet and mixed trilobite state. We see that both the transition probability and the lifetimes are slightly shorter than those computed in the spinless model. The difference can be attributed to the sensitivity of the diabaticity condition to changes in the $^3P_J$ phase shifts. The spin-less model utilizes a spin-orbit averaged $^3P$ phase shift (see Fig.~\ref{fig:sm_phases}), whose crossing position with the $^3S_1$ phase shift does not precisely mimic that of $^3P_{0,1,2}$.
However, the underlying non-adiabatic physics remains the same with/without the added spin-structure.
The Landau-Zener crossing probability is still $\gtrsim$ {0.987} and the associated lifetimes $\sim$ $\mu$s, vastly different from the adiabatic regime exhibited by $n$=27. Hence the single channel diabatic approximation for $n$=25 retains its validity whether or not the spin-structure is included.

\end{document}